\documentstyle[aaspp4,12pt]{article}
\def\Halpha{H$\alpha$}
\def\etal{{\it et al.}}
\def\IRAS{{\it IRAS}}

\def\IRDI{{\it IRAS/DIRBE}}

\def\CODI{{\it COBE/DIRBE}}
\def\ROSAT{{\it ROSAT}}
\def\mic100{\mbox{100$\mu$m}}
\def\F100{\mbox{F$_{\nu}$(100$\mu$m)}}
\def\NHoo{{\rm N_{\rm H}}}
\def\NHxo{{\rm N_{\rm H,x}}}
\def\NHIo{{\rm N_{\rm HI}}}
\def\NH21{{\rm N_{\rm H,21cm}}}
\def\NHII{{\rm N_{\rm HII}}}
\def\acm2{cm$^{-2}$}

\def\H2{H$_{2}$}

\def\pmg{$\pm$\phn}
\def\pmh{\phn$\pm$\phn}
\def\pmi{\phn\phn$\pm$\phn}

\begin{document}
\newcommand{\ARAA}[2]{\it Ann. Rev. Astron. Astrophys., \bf#1\rm, #2.}
\newcommand{\ApJ}[2]{\it Ap. J., \bf#1\rm, #2.}
\newcommand{\ApJL}[2]{\it Ap. J. Lett., \bf#1\rm, #2.}
\newcommand{\ApJSS}[2]{\it Ap. J. Supp. Ser., \bf#1\rm, #2.}
\newcommand{\AandA}[2]{\it Astron. Astrophys., \bf#1\rm, #2.}
\newcommand{\AJ}[2]{\it Astron. J., \bf#1\rm, #2.}
\newcommand{\BAAS}[2]{\it Bull. Amer. Ast. Soc., \bf#1\rm, #2.}
\newcommand{\ASP}[2]{\it Astron. Soc. Pac. Conf. Ser., \bf#1\rm, #2.}
\newcommand{\JCP}[2]{\it J. Comp. Phys., \bf#1\rm, #2.}
\newcommand{\MNRAS}[2]{\it M. N. R. A. S., \bf#1\rm, #2.}
\newcommand{\N}[2]{\it Nature, \bf#1\rm, #2.}
\newcommand{\PASJ}[2]{\it Publ. Astron. Soc. Jap., \bf#1\rm, #2.}
\newcommand{\RPP}[2]{\it Rep. Prog. Phys., \bf#1\rm, #2.}
\newcommand{\tenup}[1]{\times 10^{#1}}

\title{Measuring Molecular, Neutral Atomic, and Warm Ionized Galactic Gas \
Through X-Ray Absorption}
\author{John S.\ Arabadjis and Joel N.\ Bregman}
\affil{Dept.\ of Astronomy, University of Michigan \\
Ann Arbor, MI 48109-1090 \\
jsa@astro.lsa.umich.edu, jbregman@umich.edu}

\begin{abstract}

We study the column densities of neutral atomic, molecular, and warm
ionized Galactic gas through their continuous absorption of extragalactic
X-ray spectra at $|b| > 25^{\circ}$.  For $\NH21 < 5\tenup{20}$ \acm2,
there is an extremely tight relationship between $\NH21$ and the X-ray
absorption column, $\NHxo$, with a mean ratio along 26 lines of sight
of $\NHxo/\NH21  = 0.972 \pm 0.022$.  This is significantly less than
the anticpated ratio of 1.23, which would occur if He were half He I and
half He II in the warm ionized component.  We suggest that the ionized
component out of the plane is highly ionized, with He being mainly He II
and He III.  In the limiting case that H is entirely HI, we place an
upper limit on the He abundance in the ISM of He/H$\leq 0.103$.

At column densities $\NHxo > 5\tenup{20}$ \acm2, which occurs at
our lower latitudes, the X-ray absorption column $\NHxo$ is
nearly double $\NH21$.  This excess column cannot be due to the
warm ionized component, even if He were entirely He I, so it must
be due to a molecular component.  This result implies that for
lines of sight out of the plane with $|b| \sim 30^{\circ}$, molecular
gas is common and with a column density comprable to $\NH21$.

This work bears upon the far infrared background, since a warm
ionized component, anticorrelated with $\NH21$, might produce
such a background.  Not only is such an anticorrelation absent,
but if the dust is destroyed in the warm ionized gas, the far
infrared background may be slightly larger than that deduced by
Puget \etal\ (1996).

\end{abstract}

\section{Introduction}

The three primary mass components of the Galactic interstellar medium are
molecular gas, neutral atomic gas, and warm ionized gas.  Of these components,
the only one whose mass can be measured directly from emission is the
neutral atomic gas because the intensity of the 21 cm line of HI is directly
proportional to the HI column, barring opacity corrections, which are
generally small for sight lines out of the Galactic plane.  The mass of
molecular gas is generally determined by measuring the intensity of the
opaque CO(1$\rightarrow $0) line and applying a correction factor to convert
it to a \H2\ column.  The weakness in determining the molecular mass is
that the correction factor is not a constant of nature, but can depend upon
environment.  For the other important constituent, the warm ionized gas, one
generally obtains an emission measure (a dispersion measure is obtained
toward some pulsars), so in general, a filling factor must be assumed in
order to convert the emission measure to a column density.  Additional
information about the warm ionized gas is gained from pulsar dispersion
measures, which provide electron column densities toward individual pulsars.
Unfortunately, the dispersion measure contains no information about the
ionization state of the gas.  Due to the difficulties of obtaining the mass
of the molecular and warm ionized components, it would be valuable if there
were a simple independent approach to this issue.  Fortunately, such an
independent approach is provided by the absorption of X-rays.

Measurements of the X-ray absorption column provides a linear measure of the
mass along the line of sight, whether it be molecular, neutral atomic, or
warm ionized gas.  The reason for this is that the X-ray absorption is mainly
determined by the column density of He, which does not form compounds, or
through inner shell absorption by metals, whose cross section is nearly
independent of whether it is molecular, in dust, or free-floating.  For
neutral atomic conditions, the cross section is dominated by H and He at
energies below 0.28 keV, with He accounting for about 70-75\% of the
absorption in the 0.1-0.28 keV band.  Metal absorption begins to contribute
in the 0.28-0.5 keV range, contributing about 17\% of the total cross
section due to the presence of C and N.  Since carbon is used in some
instrumental windows, such as on \ROSAT, there is poor instrumental
sensitivity in this region.  At energies above the oxygen edge at 0.53 keV,
the metals account for at least half of the absorption, a trend that
continues to higher energies.  The absorption by metals takes place with
inner shell electrons, so the cross sections are largely independent of the
structure of the valence electrons or the degree of ionization (for modest
ionization).  Furthermore, the cross section is nearly independent of the
fraction of metals in grains, since the grains are transparent to X-rays.

Warm ionized gas has a total column density that is about one-third of the
neutral hydrogen column (Reynolds \markcite{rey89}1989, \markcite{rey91}1991),
and although hydrogen absorption is absent, helium is still a strong absorber
since it is expected to be mainly He I or He II.  The absorbing properties of
He depend on its ionization state, with He II being a less effective absorber
than He I.  Assuming that helium is half He I and half He II (Domg\"{o}rgen and
Mathis \markcite{dm}1994), the total absorption cross section is estimated
to be about 60\% that of the neutral gas near 0.2 keV.  This leads to the
expectation that absorption by warm ionized gas should be evident in X-ray
absorption measurements.

For lines of sight through total hydrogen columns of $3\tenup{20}$ \acm2,
which is typical of sight lines out of the Galactic plane, the X-ray
absorption is mainly determined by the helium column.  For total hydrogen
columns of $1\tenup{21}$ \acm2, such as occurs in sight lines through
molecular clouds or at low Galactic latitudes ($|b|<20^{\circ}$), the X-ray
absorption is mainly determined by the metals, with oxygen being especially
important.  Consequently, with suitably chosen background sources, one can
use X-ray absorption columns to study the structure of the ISM, the relative
columns of the different phases, the He abundance, the metal abundance, and
the conversion between the CO intensity and H$_{2}$.  This is the first of
several investigations by us in which we examine the relative amounts of
HI and HII as well as the He abundance in the Solar neighborhood.  This
work has an important bearing on the nature of the reported extragalactic
far infrared background, since it helps to determine the amount of far
infrared emission that could originate in warm ionized gas in the Galaxy.

This study is possible due to the ability to determine accurate X-ray
absorption columns for comparison with accurate HI columns.  The X-ray
observations discussed here are obtained with the \ROSAT\ Position Sensitive
Proportional Counter (henceforth, PSPC), which is sensitive from
0.1-2.4 keV.  Effectively, energies below 0.15 keV do not enter into the
analysis because an HI column of only $5\tenup{19}$ \acm2\ reaches
optical depth unity at 0.15 keV, and nearly all lines of sight out of the
Galaxy are greater than $1\tenup{20}$ \acm2.  X-ray absorption columns
can be determined from PSPC data with only a 5-10\% uncertainty by using
mono-temperature regions of galaxy clusters.  HI columns are taken from the
new survey by Hartmann and Burton (\markcite{hb}1997), which has an accuracy
of 5\% or less, as discussed below.  In the following sections, we describe
our assessment of the accuracy of the HI surveys (\S\ref{accHI}) and the
approach needed to obtain the most accurate X-ray absorption columns
(\S\ref{suitX}).  We apply this to determine the amount of warm ionized gas
in the Solar neighborhood, which leads to interesting constraints for the He
abundance (\S\ref{clussel},~\ref{implic}).  Also, these results
indicate that far infrared emission from dust in warm ionized gas cannot
explain the far infrared residual, which is likely to be extragalactic in
origin (\S\ref{implic}).

\section{The Accuracy of the HI Measurements \label{accHI}}

The determination of the total HI column out of the plane of the Galaxy has
been determined in several surveys, two of which minimize the effects of
sidelobe contamination: the Bell Labs survey (Stark \etal\
\markcite{sgwblhh}1992) and the recent Dwingeloo survey (Hartmann and Burton
1997, henceforth HB).  The Bell Lab survey employed a horn antenna that is
nearly free from sidelobe contamination but has only modest resolution
($2^{\circ} \times 3^{\circ}$) so that small-scale variations in the HI
column are unresolved.  The survey covered all declination regions north of
$-40^{\circ}$ and with no gaps between the beams.  The Dwingeloo survey
has better resolution (30$^{\prime}$) but it is a traditional dish design,
so all observations needed to be corrected for contamination from stray
radiation, usually by Galactic plane emission entering through the sidelobes.
This survey observed all regions north of -30 degrees declination with a beam
spacing of 30$^{\prime}$\ and a beamsize of 36$^{\prime}$, so the sky is
slightly undersampled.  The Bell Labs survey has been used widely in the past
for determinations of the absorption by the Galaxy, such as for soft X-ray
observations.  We have tried to assess whether the new Dwingeloo survey adds
a substantial improvement in accuracy for the determination of the HI column
density.

To access the accuracy of the two surveys along lines of sight out of the
Galactic plane, we have sought to compare these surveys to more accurate
measurements that are not part of an all-sky survey.  Along these lines, an
effort was made by Lockman and collaborators using the 140$^{\prime}$\
telescope to determine $\NHoo$, corrected for sidelobe contamination,
toward a variety of quasars being used in the HST absorption line key
project (Elvis, Wilkes, and Lockman \markcite{ewl}1989; Lockman and Savage
\markcite{ls}1995).  The 140$^{\prime}$\ telescope has better resolution
(21$^{\prime}$) than either the Dwingeloo or Bell Labs telescopes and
sidelobe contamination was corrected in a procedure that maps the Bell Labs
main beam, so the correction is tied to the Bell Labs measurements.  The
$\NHoo$ data from these three sources were compared for 60 lines of
sight that are representative of the full range of Galactic latitude and
longitude represented by the HB survey. Because these lines of sight were
chosen to be out of the plane, only two lines of sight have
$\NHoo > 1\tenup{21}$ \acm2\ and the central two quartiles are
bounded by the $\NHoo$ values $2.0\tenup{20}$ and $3.8\tenup{20}$
\acm2.

The internal accuracy of each survey has been estimated by a variety of
tests.  Lockman, Jahoda, and McCammon \markcite{ljm}(1986) have examined the
Bell Labs survey as it pertains to the region in directions of low HI column
density, and they estimate that the uncertainty is $1\tenup{19}$ \acm2\
($1\sigma$) when all sources of error are included.  There are several
sources of error that contribute to this figure: the accuracy of the
absolute calibration of the Bell Labs data, about 2\% (Kuntz and Danly
\markcite{kd}1992); baseline uncertainties, leading to errors of typically
$2\tenup{18}$ \acm2\ (Lockman, Jahoda, and McCammon 1986); stray radiation
from the far sidelobes give errors of about $1-5\tenup{18}$ \acm2; and noise
contributes an uncertainty of about $1-3\tenup{18}$ \acm2.  Toward the
southern limit of the survey, sidelobe contamination issues become more
difficult to correct and the accuracy is expected to be lower.  Finally, in
directions of higher column density, opacity corrections will become
important, for which there is not simple correction, and molecular gas is
likely to be present.

The uncertainties in the HB survey have been examined in detail, with
considerable attention paid to the calibration, and the near and the far
sidelobe stray radiation corrections (Hartmann and Burton 1997).  Their
accuracy is typically about $0.5-1\tenup{19}$ \acm2, or about 3\% of the
signal for lines of sight out of the plane.

Another approach to estimating the uncertainties is by comparing the results
of the different surveys in the directions observed by Lockman, Jahoda, and
McCammon (1986; henceforth LJM).  We find a smaller fractional differences
between the LJM and HB surveys than between the Bell Labs survey and either
LJM or HB.  The fractional differences between the LJM and HB survey are not
dependent upon the column density, although the most discrepant line of sight
occurs at low $\NHoo$ (Fig.\ \ref{Fig_LS}).  This particular line of sight
is at the very southern limit of the HB survey, so the sidelobe correction may
not be as accurate as at higher declinations.  Of the next three greatest
fractional differences, one is at a southern declination (-20$^{\circ}$) and
the other two are at high declination (70$^{\circ}$).  It is unclear to
us whether this is of significance since other lines of sight at such
declination values show small differences between the two surveys and these
three points are only slightly deviant from a Gaussian distribution based
upon the remainder of the population.

We examined whether the difference in $\NHoo$ depended upon the distance from
the center of the pointings of the HB survey (e.g., on the degree of the
interpolation required between pointings).  One expects a 3-4\% error to be
introduced for offsets of 10-20$^{\prime}$, based upon the structure function
analysis toward low $\NHoo$ regions (Lockman \etal\ 1986; also similar results
occur if one uses the relatively high $\NHoo$ region at $b=+6^{\circ}$; after
Bregman, Kelson, and Ashe \markcite{bka}1993).  This error introduced by the
offset is significantly smaller than the 7.8\% rms and was not easily isolated
in our sample, but is an effect that is probably present.  In the comparison
of the two surveys, a slight concern is that there is a weak anticorrelation
between the fractional $\NHoo$ difference and $\NHoo$ (Fig.\ \ref{Fig_LS}).
Also there is a slight offset in the mean $\NHoo$ between the two works, but
it is only a $1.7\sigma$ effect.

One may assume that the absolute error in the HB and Lockman surveys are the
same and that there is a 3-4\% difference in $\NHoo$ introduced by the need
to interpolate in the HB survey.  Under these assumptions, we find that the
true uncertainty of a measurement point in each sample is 5.0\%.  The error
in the HB measurements are probably closer to 3\% (discussed in Hartmann and
Burton 1997), with 5\% being a conservative value.

The accuracy of determining $\NHoo$ in a line of sight using the Bell
Labs data can be estimated in the same fashion, although in this case, the
result from the Bell Labs data can be compared to both the HB survey and the
Lockman columns.  The comparison between the Bell Labs data and the Lockman
data should have the greatest internal consistency since Lockman anchors
their measurements to the Bell Labs observations.  The standard deviation
between the two sets of measurements is 9.5\% and the mean offset is 0.031, a
$2.5\sigma$ difference from zero.  Most of the difference between the two
measurements can be attributed to the variation within the rather large beam
of the Crawford Hill telescope.  In the comparison with the HB survey, the
offset is 0.054, a $3.5\sigma$ difference and the standard deviation is
12\%.  This difference and standard deviation is larger than anticipated.

The typical angular scale over which our X-ray column densities are
determined is about $12^{\prime}$, a factor of three smaller than the
$36^{\prime}$ diameter beam of the Dwingeloo observations, and so we sought
to characterize the magnitude of the expected emission fluctuations within
the Dwingeloo beam with an independent method.  This was accomplished by
analyzing the power spectrum of the 21 cm emission.  The spatial power
spectrum of HI structure is obtained by squaring the Fourier transform of a
21 cm sky brightness map (e.g. Crovisier and Dickey \markcite{cd}1983).  The
integral of the power spectrum, weighted by the square of the
Fourier-transformed windowing function, results in a measure of the mean
square emission contrast of the observed region (see, e.g., the discussion
in Peebles \markcite{peeb}1993).  Using the power spectrum obtained by
Crovisier and Dickey (1983) for Arecibo observations of HI in the Galactic
plane, we calculated an rms emission contrast of 7\% using a
{\sc fwhm}=30$^{\prime}$ Gaussian window, consistent with our previous
estimate.

In summary, the HB survey provides the best measurements of $\NHoo$ over most
of the sky, and for directions out of the plane and where opacity effects are
unimportant, the uncertainty in a measurement should be 5\%, if not better.
In the comparison between X-ray absorption columns and HI columns, we will
concentrate on regions where $\NHoo < 3\tenup{20}$ \acm2\ and away from the
southern limit of the HB survey.

\section{Suitability of X-Ray Sources for Absorption Studies
\label{suitX}}

The goal of this X-ray analysis is to obtain X-ray absorption column
densities through the Galaxy with uncertainties near 10\%, which is needed
to distinguish between absorption by warm ionized gas and by neutral gas. To
achieve this goal, we examined the accuracy to which absorption columns
could be determined for three classes of extragalactic sources: active
galactic nuclei; early-type galaxies, and galaxy clusters.  The most common
background sources are AGNs, but there were two problems in determining
accurate column densities.  First, the intrinsic underlying spectrum is not
known and a power-law description may be inadequate.  Second, and possibly
more important is that many AGNs exhibit intrinsic absorption of their own,
so separating this from absorption by the Galaxy is difficult.  Our fits for
the X-ray absorption column toward several AGNs led to a mean value of
$\NHoo$ that was about 10-20\% above the 21 cm value of $\NHoo$, which is
about the difference expected by the absorption from warm ionized gas (see
also Fiore \etal\ \markcite{femsw}1994).  However, we could never be
confident that this difference was not due the two problems stated above, so
we searched for more suitable background sources.

We considered early-type galaxies as potential background sources since
their spectrum can often be characterized by a single temperature thermal
plasma.  For this fit, one needs to assume that a single-temperature is an
accurate model and a metallicity must be assumed or fit to the spectrum.
Unfortunately, there is considerable debate as to the correct metallicity
for these systems, and the value of the metallicity assumed has a direct
influence on the value of $\NHoo$ that one obtains from the X-ray fits.
Furthermore, multiple temperature models are needed for at least some galaxies
(e.g., NGC 4365, Fabbiano \etal\ \markcite{fkt}1994; Buote and Fabian 1998),
and there is some concern about the accuracy of the thermal plasma models at
temperatures of 0.1-1 keV (Bauer and Bregman \markcite{bb}1996). Despite
these problems, we examined the X-ray absorption columns toward 16 of the
brightest X-ray emitting early-type galaxies.  The uncertainty in these
columns were too large to make these useful sources, and often, an extremely
low X-ray absorption column was indicated (well below $\NHoo$ from 21
cm observations), which is non-physical.  This latter result could be
understood if there is some X-ray emitting gas below the mean temperature,
which would be expected from cooling flows.  However, until we have adequate
information to create accurate and well-understood cooling flow models for
these galaxies, these sources are poor choices to obtain the X-ray absorbing
columns in this study.

Galaxy clusters appear to be the most useful targets against which to
determine the X-ray absorbing columns because there are regions in clusters
that provide simple well-defined spectra.  The spectrum of galaxy clusters
may be complex in the center due to cooling flows, but this is confined to
within the cooling radius, about 100-200 kpc, or 1-2$^{\prime}$\ for a
cluster at z = 0.05.  Beyond the central region, clusters appear to be fairly
isothermal until the very outer parts near 1-2 Mpc, although these outer
regions have a low surface brightness.  Consequently, the region beyond the
cooling flow, but where the X-ray surface brightness is still high, is an
ideal region in which to define a spectrum.  The gas temperature is
characteristic of the mean cluster temperature and the mean metallicity has
been measured, both by independent instruments (Einstein, GINGA, and ASCA).
The temperature of the cluster gas is typically 5 keV, so the spectrum is
dominated by free-free radiation, with the line contribution coming primarily
from He-like and H-like ions, which are the easiest and most reliable to
model. Furthermore, uncertainties in the gas temperature and metallicity have
little effect in the determination of the X-ray absorption column because the
metals are only weak contributors and because there is practically no
difference in the spectrum within the \ROSAT\ PSPC band of a 5 keV
spectrum compared to a 6 keV or 4 keV spectrum, which is typical of the
temperature uncertainties for a cluster.

\section{Cluster Selection and the Resulting X-ray Absorption Columns
\label{clussel}}

The types of clusters that were selected for this study fulfill a variety of
properties, based upon Galactic latitude and longitude, X-ray brightness,
and whether basic parameters are known for the hot gas, such as temperature
and metallicity.  The X-ray brightness must be sufficient such that the
spectrum in an annulus outside the center has enough counts to place tight
constraints on spectral fits.  In obtaining spectra, it is advantageous to
determine the background in the same exposure as the cluster, so we avoided
the nearby and large Virgo cluster.  It is desirable for the X-ray
temperature and metallicity to be known so that the fitted PSPC spectrum has
the fewest possible free parameters.  Finally, the emphasis of the study is
on modest and low Galactic column densities, typically $< 5\tenup{20}$ \acm2,
because for higher columns, molecular gas starts to become more common and
opacity corrections must be considered for the neutral hydrogen column.
Given the desired cluster characteristics, we began the study by choosing
clusters from the flux limited high galactic latitude sample of Henry and
Arnaud \markcite{ha}(1991; $F_x($2-10 keV$) > 3\tenup{-11}$ erg \acm2\
s$^{-1}$, $|b| > 20^{\circ}$) with the addition of the clusters Hydra-A and
A2163, which were later found to meet the criteria.  This is a complete
sample, which is not absolutely necessary for a study such as this, since
the modest Galactic columns do not bias the hard flux from which the sample
was chosen.

Each of the clusters that was studied (Table \ref{tab:clustlist}) was
processed with both the current XSPEC (see Arnaud \markcite{arn}1996
for a review) and PROS (e.g. Conroy \etal\ \markcite{cdmors}1993)
packages with standard techniques.  Spatial and temporal variations in the
instrument (\ROSAT\ PSPC B or C; see, e.g., Briel, Burkert, and
Pfeffermann \markcite{bbp}1989) were corrected for using the latest
PCPICOR task in FTOOLS 4.0, before spectra were extracted using
the most recent response matrices (for PSPC B gain 1-2 or PSPC C
gain 1).  The data were screened for periods of high background and a
background-subtracted spectral energy distribution was produced in a pair of
annuli that are usually 3-6$^{\prime}$ and 6-9$^{\prime}$, with point sources
removed.  Exceptions to this include A0665 (3-6$^{\prime}$, 9-12$^{\prime}$),
A2163 (2-4$^{\prime}$, 4-6$^{\prime}$), A2256 (6-9$^{\prime}$,
9-15$^{\prime}$), and A1656 and A2199 (two off-center circular regions), where
we wished to avoid multiple point sources or galaxies, or modulate the number
of counts.  Each spectrum was fit with a single-temperature model whereby the
metallicity and temperature are fixed, and taken from other sources (White,
Jones, and Forman \markcite{wjf}1997; White \etal\ \markcite{wfjma}1991).  By
fixing these parameters, there are only two remaining parameters in the
spectral fit: the normalization and the absorption column.  

The most recent absorption correction procedure available in XSPEC is
based upon the work of Ba{\l}uci{\'{n}}ska-Church and McCammon
\markcite{bcm}(1992; in the routine VPHABS), who include autoionization
corrections for He as well as a cross section that is different from tha
absorption model of Morrison and McCammon \markcite{mm}(1983; in the
routine WABS).  However, recent work by Yan, Sadeghpour, and Dalgarno
(\markcite{yan}1998), who calculate a more accurate He cross section, shows
that the He cross section in the soft X-ray band given by
Ba{\l}uci{\'{n}}ska-Church and McCammon \markcite{bcm}(1992) is too great.
Consequently, we wrote another absorption routine for use in XSPEC that
incorporates the accurate He cross section of Yan \etal\ (1998).  All of
our fitting is performed with this higher accuracy absorption routine.  For
the purposes of this paper, one of the important parameters is found to be
the He to H ratio, and we generally use the fairly standard value of 0.10
(e.g., Osterbrock, Tran, and Veilleux 1992), but we also consider a value
of 0.09, which was deduced from observations of the Orion nebula by
Baldwin \etal\ (1991).

A successful fit is one in which the best-fit has at least a 5\% chance of
occurrence based upon the $\chi^2$ (e.g., for 187 degrees of freedom, the
reduced $\chi^2 < 1.18$, or 1.26 for 1\%).  In a few clusters the number
of counts in each channel was sufficiently large that errors in the PSPC
calibration produce an artificially large contribution to the $\chi^2$.  In
those cases we replaced our central source annuli with regions at larger
distance and lower flux, producing fits with acceptable $\chi^2$ values.
Spectral fits to A1795 and 2A0335 are shown in figures \ref{Fig_A1795fit}
and \ref{Fig_2A0335fit} to demonstrate the wide range in Galactic absorption.
Uncertainties in the resulting X-ray absorption column are based upon the
deviation from the best-fit, which is the usual procedure.  We show confidence
contours in the two fitting parameters for A1795 and 2A0335 in Figures
\ref{Fig_A1795con} and \ref{Fig_2A0335con}.  The resulting accuracy of the
X-ray absorption column, expressed as $\NHxo$ is typically 5-10\%, as seen
in Table \ref{tab:mods}.  We include the free electron column density of the
Taylor and Cordes \markcite{tc}(1993) model for comparison therein.  The
X-ray and 21 cm HB columns are compared in Figure \ref{Fig_NHxNH}.

\section{The Nature of Warm Ionized and Molecular Gas \label{implic}}

There is an astonishing division in the nature of the X-ray absorption near
$\NHxo$ = $5\tenup{20}$ \acm2 (Fig. \ref{Fig_NHxNH}; Table \ref{tab:mods}).
For X-ray absorption columns below that value, the neutral hydrogen column
entirely accounts for the X-ray absorption.  However, for larger X-ray
columns, there must be a gaseous absorption component in addition to the
neutral hydrogen.  Because of this natural division, we will discuss the
implications of the results in these two separate regimes.  These results
have important consequences for the nature of the thick warm ionized
``Reynolds'' layer in the Galaxy, for the nature of molecular gas, as well
as for the determination of the extragalactic far infrared background.

\subsection{Implications for the Galactic Warm Ionized Layer \label{warmion}}

As discussed above, the X-ray absorption column should reflect the
contributions from molecular gas, neutral gas, and warm ionized gas (provided
most of the He is either He I or He II).  Consequently, it was surprising to
find that for $\NHxo < 5\tenup{20}$ \acm2, there is little room for
contributions from any component besides neutral atomic gas.  In order to
quantify this result, we introduce the quantity $C$, defined through the
expression $\NHxo = C \cdot \NH21$.  For $\NHxo < 5\tenup{20}$ \acm2,
$C = 0.972 \pm 0.022$ (this is the average; the error-weighted average is
nearly identical).  Our statistical tests indicate that there is no
significant trend of $C$ with column density and that the standard deviation
for a single point is 11\%.  The uncertainty in $\NH21$ was estimated to be
5\%, so most of the contribution to the standard deviation is inferred to lie
with $\NHxo$ (about 10\%; this is slightly larger than the median error
associated with the X-ray fits).  When we include the uncertainty in $\NH21$
with the uncertainty in $\NHxo$ from the spectral fit, there is only one point
that lies more than 2.5 sigma below the average line: the outer region of the
Coma cluster.  This may reflect unresolved variation in $\NH21$ by radio
telescopes or the difficulty of correcting for stray radio radiation at the
lowest column density levels.

Our results indicate that, at the 99\% confidence level, $C < 1.023$, which
means that only 2.3\% of the optical depth could be caused by some component
other than neutral hydrogen.  At these column densities, the amount of
molecular hydrogen is found to be small (Savage \etal\ 1977) and would not be
detectable as excess X-ray absorbing material.  However, the warm ionized
component of interstellar gas was expected to be detectable at somewhat
greater levels.

Warm ionized gas has been studied most extensively by Reynolds (1989, 1991),
who finds that in the Solar circle, the integrated HII column through the
entire disk is $2.3\tenup{20}$ \acm2, while the total HI column is
$6.2\tenup{20}$ \acm2.  Therefore, the HI+HII column is 37\% larger than the
HI column alone.  However, $\NHxo$ is not 37\% larger than $\NH21$ because
the opacity is lower in the warm ionized gas.  If HeII/He = 0.5 in the warm
ionized gas (Domg\"{o}rgen and Mathis 1994; see discussion by Snowden \etal\
\markcite{shjlms}1994), the warm ionized gas has an effective cross section
that is about 62\% of neutral gas at about 1/4 keV.  This leads to the
prediction that $C$ = 1.23, which is significantly above the value of $C$
found from our data.

There are a few possible explanations for the difference between the
anticipated and observed ratio of $\NHxo/\NH21$.  First, it is important to
recognize that the studies of \Halpha\ emission from diffuse gas relates to
the emission measure and can be converted to a column density only after
adopting a filling factor for the gas.  The determination of the electron
column density derives directly from the pulsar dispersion measure (Taylor
and Cordes 1993), although a dispersion measure provides no information
about the ionization state of the gas.  By combining these two types of
measurement, filling factors can be determined, upon assuming that the gas
causing the dispersion measure and the \Halpha\ emission are one and the same.

One possible explanation is that our sight lines happen to avoid most
of the ionized halo gas that causes the pulsar dispersion measure.  The 
ionized gas out of the plane has a scale height of about 1 kpc, so Taylor
and Cordes (1993) quantify this distribution by measuring the dispersion
measure toward pulsars in globular clusters and the scattering measure from
AGNs.  Based upon the observations of globular cluster pulsars, there do not
appear to be frequent holes in the electron distribution, so it is unlikely
that our 13 galaxy clusters lie in regions of low electron column densities.

Another possibility is that the warm ionized gas is more highly ionized than
previously anticipated, thereby reducing the X-ray absorption cross section.
Calculations of the ionization state of the warm ionized gas place the HeI/He
fraction at about 50\% (Domg\"{o}rgen and Mathis 1994) and we arrive at a
similar result (for the mean ISM ionizing spectrum and CLOUDY; Ferland
\markcite{fer}1996).  As mentioned above, this results in $C=1.23$, far above
our upper limit of 1.023.  The situation is only marginally improved if all
of the He is singly ionized; in that case $C=1.18$.  In order to be consistent
with our X-ray absorption measurements, He would have to be at least 87\%
HeIII.  Lowering the value of He/H to 0.09 reduces the required ionization
fraction by less than 1\%.

A high degree of ionization could also explain the lack of recombination line
radiation from He I $\lambda5876$ (Reynolds and Tufte \markcite{rt}1995).
For gas in photoionization equilibrium, this would require at least an order
of magnitude increase in the density of ionizing radiation than estimated to
exist in the Galactic disk and would require that there be an additional
source of ionization.  Alternatively, the electrons may be associated with
cooling gas in a galactic fountain, which would naturally have a high degree
of ionizionation.  In order for this gas not to exceed the disk pressure, the
mean temperature of the cooling gas would need to be no greater than about
$5\tenup{5}$ K.

Another factor that enters into this analysis is the value of the He/H
ratio.  We have used a He/H ratio of 0.10 for the above analysis, which is
the common value used in X-ray astronomy and is comparable to the Solar
value of 0.098.  However, the value in the interstellar medium has been
determined from the HII regions in the Orion nebula by two groups, who find
ratios of 0.089 and 0.101, with the differences being attributed to details
about the dust properties (Baldwin \etal\ \markcite{bfmccps}1991; Osterbrock
\etal\ 1992).  If this lower value is correct, then $C$ increases by 0.07 to
1.046 (Fig.\ \ref{Fig_HHe}), and at the 99\% confidence level, $C < 1.10$.
This limit on $C$ is reached if at least 45\% of the He is He III, with the
remainder being He II.

We suggest that the most likely explanation for the low value of $C$ is that
the gas is more highly ionized than gas in a typical hot HII region.  The
gas may be hot galactic fountain gas at $10^5$-$10^6$ K or cooler gas that is
subject to a more intense ionizing radiation field than previously believed.
For this explanation to be sensible, the \Halpha\ emitting gas would not be
cospatial with the dispersion measure gas.  Consequently, the \Halpha\
emission would need to originate from denser, lower filling factor material
than assumed by Reynolds (1989, 1991).

In passing, we note that our observations can be used to set an upper limit
to the He abundance in the ISM by requiring that the X-ray absorption column
not be smaller than the 21 cm column (i.e., $C \geq 1$).  If we demand that
$C = 1$ ($\NHxo = \NHIo$, and $\NHII = 0$) then, at the 99\% confidence level,
He/H $\leq$ 0.103.

\subsection{The Detection of Molecular Gas by X-Ray Observations \label{molec}}

One of the striking results for our sample is that for $\NHxo > 6\tenup{20}$
\acm2, the amount of X-ray absorbing material is substantially in excess
of the measured atomic column.  Seven clusters show this significant excess
absorption, which requires an equivalent neutral atomic absorption that is
comparable to the 21 cm HI column.  This is over an order of magnitude greater
than the amount expected from warm ionized gas in these directions (Taylor and
Cordes 1993), even if all of the He were in the form of He I.  Consequently,
the only viable candidate for this excess absorption is molecular gas.  In
support of this suggestion, we note that the presence of this extra
absorption occurs at the same column density where molecular gas can
self-shield against ultraviolet photons (Federman, Glassgold, and Kwan
\markcite{fgk}1979) as well as at the column above which \H2\ becomes
abundant, based on Copernicus ultraviolet absorption line observations toward
stars (Savage \etal\ 1977).

In order to extract the molecular column density from these measurements, we
correct for the small opacity in the 21 cm line (3\% effect) and for the
slight increase in the X-ray opacity because \H2\ has a cross section
that is 40\% larger, per H atom than HI (Yan, Sadeghpour, and Dalgarno
1998).  At these column densities, about half of the opacity is due to \H2\
and He (with He being the dominant of the two) with the remainder due to O,
C, and N (O is the dominant metal absorber).  We assume Solar abundances
for the metals and that the X-ray absorption cross section for the metals is
independent of whether they are in molecular or atomic form.  The resulting
fractional \H2\ column is compared to the total inferred column of atoms,
N(HI+2H$_2$) (Fig.\ \ref{Fig_H2frac}), similar to the approach taken by
Savage \etal\ (1977), whose results from ultraviolet \H2\ studies are also
shown.  Our primary result is that for sightlines out of the disk where
$\NHxo > 6\tenup{20}$ \acm2, the amounts of atomic and molecular hydrogen
are comparable.  Our mean molecular fraction is about twice that of Savage
\etal, for the same range of total atomic columns.

There are a few important differences in the location and distance of targets
between our sample and that of Savage \etal\ \,  Six of the seven clusters
lie in the range $26^{\circ}< |b| < 40^{\circ}$ (the seventh lies at
$b = -50^{\circ})$ and the sight lines sample the entire vertical extent of
the disk.  In the Savage \etal\ sample, all of the stars more distant than
0.5 kpc lie either at very low latitude in the plane, or at fairly high
latitude where N(H$_2$) is quite low.  There are no distant stars at
$|b| = 25-40^{\circ}$, and our sample has no clusters at $|b| < 25^{\circ}$.
Our work samples the atomic and molecular gas in the Solar neighborhood at
mid-latitudes, while the Savage \etal\ work primarily samples sight lines
through the disk and this difference in sampling may be responsible for the
slightly different molecular gas fractions.  Also, the stars in Savage
\etal\ are hot stars, which may be able to destroy local molecular gas,
thereby decreasing the molecular fraction, an effect that would be
greatest for the nearest stars.

Another investigation that our work can be compared to is the study of
HCO$^+$ and OH absorption toward extragalactic sight lines (Lucas and Liszt
\markcite{luclis}1996 and references therein).  At Galactic latitudes below
$39^{\circ}$, 18 of 22 sightlines were detected as HCO$^+$ absorbers,
indicating that molecular gas is quite common.  They must adopt a conversion
of N(HCO$^+$) to N(H$_2$), so there is greater uncertainty than the direct
measurement of \H2\ by Savage \etal\ (1977).  Nevertheless, their molecular
hydrogen columns are similar to those of Savage \etal\ (1977), and slightly
lower than our molecular fractions.  Attempts to measure this \H2\ gas by
CO surveys at moderate and high galactic latitudes fail to detect abundant
CO emission (Hartmann, Magnani, and Thaddeus \markcite{hmt}1998), a result 
that must reflect the fact that CO is destroyed before \H2.  Finally, we
note that molecular gas may be a considerable if not dominant fraction of
the gas mass in the outer part of the Milky Way according to Lequeux, Allen,
and Guilloteau \markcite{leq}(1993).

The presence of molecular gas is reflected in the far infrared emission as
observed by the \IRAS\ satellite.  We extracted the \IRAS\ \mic100\
observations in the direction of each of the clusters, as shown in Figure
\ref{Fig_IRAS}.  Each of the lines of sight whose column density
$\NH21 > 5\tenup{20}$ \acm2\ (i.e. the filled circles in
Fig.\ \ref{Fig_NHxNH}) shows moderate or substantial structure in its
\mic100\ emission.  For example, 2A0335 lies behind a large tongue of far
infrared \IRAS\ emission and its position is adjacent to a known molecular
cloud.  It is possible that molecular gas, just below the threshold of the
instrument, lies in front of 2A0335.

Earlier, we considered the possibility that fluctuations in the HI
distribution might lead us to infer 21 cm columns that were inaccurate, but
the analysis of 21 cm data indicated that the uncertainties would be at the
5\% level in $\NHIo$.  This test can also be conducted by examining the
fluctuations in the total gas column as reflected in the \IRAS\ \mic100\
fluxes.  In particular, we examine whether there might exist large column
density fluctuations within the 36$^{\prime}$ diameter beam of the HI survey.
The most direct method of making this test is to compare the \F100\ within
the annuli used for the X-ray measurements with the \F100\ within a
36$^{\prime}$ diameter region.  Using \mic100\ images centered on the
clusters, we divided each image into 6 concentric annuli, each with a
thickness of 3$^{\prime}$.  We calculated the \mic100\ flux averaged over
each annulus (i.e. the differential mean) and also over the entire disk
contained within each annulus (the cumulative mean; Fig.\ \ref{Fig_ann}).
In each case we calculated the relative difference $\delta f_{\nu}$ between
\F100\ averaged over the 36$^{\prime}$ (diameter) disk and \F100\ averaged
over the 6-12$^{\prime}$ annulus, corresponding to the HB beam and the X-ray
absorption calculations, respectively.  The mean value of $|\delta f_{\nu}|$
for the seven clusters is 5\%, and the mean value of $\delta f_{\nu}$ is
-0.3\%, indicating that the errors in \F100\ are consistent with being
random at the 5\% level.  This is the same degree of fluctuation found from
the HI analysis.

The \mic100\ flux is known to correlate with HI gas, although nonlinearly
and with significant scatter (e.g., Boulanger \etal\ \markcite{boul}1996).
Here we examine whether the correlation is modified when correlating the
X-ray absorption column with the far infrared emission.  We obtained values
for the \mic100\ fluxes from the public \IRAS\ products (using IRSKY 2.3.1),
in which the Zodiacal light contribution has been removed with a model,
although this is not the latest model (Shafer \etal\ 1997).  A least-squares
fit between $\NHIo$ and \F100\ yields
$\NHIo \propto$ F$_{\nu}$(\mic100)$^{0.57\pm 0.29}$.  The departure of the
exponent from unity was thought to be due in part to the inability to include
N(H$_2$) along the line of sight, and to the oversubtraction of the estimated
Zodiacal flux at low levels.

Improved determinations of the far infrared fluxes are available through 
the recent work of  Schlegel \etal\ \markcite{sfd}(1998), who have
reprocessed the \IRAS\ data together with the \CODI\ maps.  Relative to
the older standard \IRAS\ products, these values for F$_{\nu}$(\mic100) are
often significantly different at low flux levels, generally being larger.
This leads to steeper relationships, with
$\NHIo \propto {\rm F}_{\nu}$(\mic100)$^{0.84\pm 0.27}$ and
$\NHxo \propto {\rm F}_{\nu}$(\mic100)$^{1.16\pm 0.22}$
(Fig.\ \ref{Fig_NHF100}), consistent with unity.

\section{Implications for the Far Infrared Background \label{FIR}}

These results have an important consequence for the intensity of the far
infrared background (Puget \etal\ 1996; Boulanger \etal\ \markcite{boul}1996)
because those investigators needed to remove the Galactic far infrared
signal associated with the neutral and ionized gas around the Sun.  After
correcting for Zodiacal emission and the 3 K background, they remove a
Galactic far infrared signal that is proportional to the HI column, resulting
in a positive residual.  This is the only correction to Galactic gas if the
relative columns of neutral and warm ionized gas are constant and if the dust
temperature in these regions is independent of the column density.  However,
there is the concern that this ratio of the neutral to warm ionized column
densities is not a constant and that the warm ionized column increases as
$\NHIo$ decreases.  These investigators consider that possibility and
subtract an additional far infrared signal that is due to an additional
column density of ionized gas of amount $4\tenup{19}$ \acm2.  In comparison,
Schlegel \etal\ (1998) and Shafer \etal\ \markcite{smfbj}(1997) also find
evidence for a far infrared background above \mic100, although they do not
explicitly remove a separate column density of warm ionized gas.

Based upon our X-ray absorption measurements, there is no evidence for an
inverse correlation between the neutral and warm ionized gas.  As shown
above, for $\NHIo < 5\tenup{20}$ \acm2, there is little room for warm
ionized gas of moderate ionization state, although we know that ionized
material is present at about the level assumed by Puget \etal\ (1996) based
upon the pulsar dispersion measurements.  If the dust is destroyed in this
ionized material, as might occur if the gas had been heated to $\sim 10^6$ K,
as in a galactic fountain picture, then Puget \etal\ (1996) have subtracted
too much far infrared emission, and the far infrared background is somewhat
brighter than they calculated.  However, if the dust is not destroyed in
this ionized column, and if it is at the mean temperature of dust nearer the
disk, then they have dealt with this contribution correctly, in which case
the far infrared background as determined by Shafer \etal\ (1997) may be
too high.

We have examined the utility of using the X-ray columns and the
F$_{\nu}$(\mic100) of Schlegel \etal\ (1998) to examine the presence of the
far infrared background.  When we perform a linear fit of F$_{\nu}$(\mic100)
on $\NHIo$, we find that when $\NHIo$ = 0,
F$_{\nu}$(\mic100)$ = 0.04 \pm 0.31$ (Fig.\ \ref{Fig_IR21}), which is
indistinguishable from zero, as Schlegel \etal\ (1998) found, for a larger
data set.  However, when we compare the X-ray column to the far infrared
flux, we obtain a background F$_{\nu}$(\mic100)$ = 0.26\pm 0.10$ MJy
sr$^{-1}$ (Fig.\ \ref{Fig_IRX}), which is slightly larger but consistent with
the value of Puget \etal\ (1996), 0.12 MJy sr$^{-1}$.  We do not claim that
our work shows a convincing detection of the far infrared background, since
it is not even a $3\sigma$ result.  However, we hope that this demonstrates
the feasibility of the approach and we note that the sample can be expanded
substantially, probably to 100-200 clusters, which would reduce the
uncertainty.  Also, future instruments will have better spectral
capabilities, which will improve the accuracy of individual measurements
of $\NHxo$.

We would like to give special thanks to F.J.\ Lockman and D.\ Hartmann for
their responses and guidance to the many questions that we raised.  Also, we
would like to thank X.\ Desert, J.P.\ Puget, Eli Dwek, Bill Reach, S.\
Snowden, D.\ McCammon, K.\ Arnaud, C.\ McKee, J.\ Cordes, S.\ Spangler, R.\
Lucas, S.\ Tufte, R.\ Reynolds, M.\ Bremer, and G.\ Mamon for stimulating
conversations on this topic.  JNB would like to thank A.\ Omont as well as
the faculty and staff at the Institut d'Astrophysique for their hospitality
during which this work was begun.  This work made extensive use of the
\ROSAT\ data archive facility at the HEASARC as well as the NASA
Extragalactic Database.  Archived infrared data were obtained with IRSKY
2.3.1 and the FITS interface and maps of of D.\ Schlegel, D.\ Finkbeiner,
and M.\ Davis.  Financial support for this work was provided for by NASA
grants NAG5-3247 and  NAG5-3352.

\vspace{20pt}

\begin{deluxetable}{rrrcl}
\tablewidth{250pt}
\tablecaption{The cluster sample. \label{tab:clustlist}}
\tablehead{
\colhead{cluster}        &
\colhead{{\it l}$^{II}$} &
\colhead{{\it b}$^{II}$} &
\colhead{T\tablenotemark{a}, keV } &
\colhead{z\tablenotemark{a} }
}
\startdata
 2A0335  & 176.25 & --35.08 & {\phn}3.1 & 0.035  \nl
  A0085  & 115.05 & --72.08 & {\phn}6.2 & 0.0521 \nl
  A0119  & 125.75 & --64.11 & {\phn}5.1 & 0.0443 \nl
  A0133  & 149.09 & --84.09 & {\phn}3.8 & 0.0604 \nl
  A0401  & 164.18 & --38.87 & {\phn}7.8 & 0.0748 \nl
  A0478  & 182.43 & --28.29 & {\phn}6.8 & 0.0881 \nl
  A0496  & 209.59 & --36.49 & {\phn}4.7 & 0.0330 \nl
  A0665  & 149.73 &  +34.67 & {\phn}8.3 & 0.1816 \nl
  A1060  & 269.63 &  +26.51 & {\phn}3.3 & 0.0124 \nl
  A1651  & 306.83 &  +58.62 & {\phn}7.0 & 0.0825 \nl
  A1656  &  58.16 &  +88.01 & {\phn}8.0 & 0.0231 \nl
  A1795  &  33.81 &  +77.18 & {\phn}5.1 & 0.0621 \nl
  A2029  &   6.49 &  +50.55 & {\phn}7.8 & 0.0765 \nl
  A2052  &   9.42 &  +50.12 & {\phn}3.4 & 0.0348 \nl
  A2142  &  44.23 &  +48.69 &      11.0 & 0.0899 \nl
  A2147  &  28.83 &  +44.50 & {\phn}4.4 & 0.0356 \nl
  A2163  &   6.75 &  +30.52 &      13.9 & 0.2030 \nl
  A2199  &  62.93 &  +43.69 & {\phn}4.7 & 0.0299 \nl
  A2256  & 111.10 &  +31.74 & {\phn}7.5 & 0.0581 \nl
  A2657  &  96.65 & --50.30 & {\phn}3.4 & 0.0414 \nl
\enddata
\tablenotetext{a}{T, z from White, Jones and Forman (1997) and references
therein, except for 2A0335, which is from White \etal\ (1991).}
\end{deluxetable}

\begin{deluxetable}{rllcl}
\tablewidth{350pt}
\tablecaption{Column densities toward each cluster in the sample (in units of
$10^{20}$ \acm2).
\label{tab:mods}}
\tablehead{
\colhead{cluster} &
\colhead{$\NHxo(1)$ \tablenotemark{a}}  &
\colhead{$\NHxo(2)$ \tablenotemark{a}}  &
\colhead{$\NH21$ \tablenotemark{b}}     &
\colhead{N$_{\rm e}$ \tablenotemark{c}}
}
\startdata
2A0335 &      27.8 \pmi 1.1   &      26.1  \pmi 1.7   &  17.4\phn  & 0.855 \nl
A0085 & \phn  2.78 \pmh 0.07  & \phn  2.79 \pmh 0.10  & {\phn}2.89 & 0.534 \nl
A0119 & \phn  3.35 \pmh 0.20  & \phn  3.10 \pmh 0.18  & {\phn}3.45 & 0.562 \nl
A0133 & \phn  1.46 \pmh 0.06  & \phn  1.29 \pmh 0.09  & {\phn}1.60 & 0.512 \nl
A0401 &      12.6  \pmi 1.0   &      15.3  \pmi 1.7   & {\phn}9.57 & 0.787 \nl
A0478 &      33.9  \pmi 0.9   &      37.4  \pmi 1.5   &  14.1\phn  & 1.02  \nl
A0496 & \phn  7.02 \pmh 0.34  & \phn  7.13 \pmh 0.52  & {\phn}4.11 & 0.830 \nl
A0665 & \phn  4.73 \pmh 0.24  & \phn  4.49 \pmh 0.88  & {\phn}4.24 & 0.867 \nl
A1060 & \phn  6.75 \pmh 0.41  & \phn  6.71 \pmh 0.33  & {\phn}4.87 & 1.15  \nl
A1651 & \phn  1.59 \pmh 0.08  & \phn  1.25 \pmh 0.11  & {\phn}1.47 & 0.602 \nl
A1656 & \phn 0.781 \pmg 0.050 & \phn 0.597 \pmg 0.071 & {\phn}0.90 & 0.509 \nl
A1795 & \phn 0.909 \pmg 0.026 & \phn 0.909 \pmg 0.017 & {\phn}1.04 & 0.525 \nl
A2029 & \phn  3.23 \pmh 0.08  & \phn  3.23 \pmh 0.13  & {\phn}3.18 & 0.673 \nl
A2052 & \phn  3.10 \pmh 0.13  & \phn  2.70 \pmh 0.18  & {\phn}2.71 & 0.676 \nl
A2142 & \phn  4.17 \pmh 0.14  & \phn  3.88 \pmh 0.21  & {\phn}3.86 & 0.688 \nl
A2147 & \phn  2.52 \pmh 0.65  & \phn  2.56 \pmh 1.12  & {\phn}2.82 & 0.741 \nl
A2163 &      22.4  \pmi 1.6   &      26.4  \pmi 2.5   &  11.0\phn  & 1.05  \nl
A2199 & \phn 0.830 \pmg 0.035 & \phn 0.877 \pmg 0.047 & {\phn}0.81 & 0.744 \nl
A2256 & \phn  4.47 \pmh 0.14  & \phn  4.65 \pmh 0.13  & {\phn}4.16 & 0.950 \nl
A2657 &       11.3 \pmi 0.8   &      14.5  \pmi 1.4   & {\phn}6.00 & 0.660 \nl
\enddata
\tablenotetext{a}{$\NHxo$ toward two different source regions in each cluster.}
\tablenotetext{b}{$\NH21$ from Hartmann and Burton (1997).}
\tablenotetext{c}{N$_{\rm e}$ calculated from the model of Taylor and Cordes
(1993).}
\end{deluxetable}

\clearpage

{\bf Figure captions}

\vspace{15pt}
\noindent Fig\_LS.ps

\figcaption{The fractional difference between the HI column densities of 
Hartmann and Burton (1997) compared to those of Lockman and collaborators
(Elvis, Wilkes and Lockman 1989; Lockman and Savage 1995) where the absicca
is $\NHIo$ from Hartmann and Burton (1997).  The lines of sight are toward
active galactic nuclei, and the fractional difference is
(N(Hartmann-Burton)-N(Lockman))/N(Lockman).  \label{Fig_LS}}

\vspace{15pt}
\noindent Fig\_A1795fit.ps

\figcaption{A spectral fit to A1795, a low-absorption column cluster.  The top
panel shows the model spectrum convolved with the PSPC instrument
response; the residuals are shown at the bottom.  \label{Fig_A1795fit}}
 
\vspace{15pt}
\noindent Fig\_2A0335fit.ps

\figcaption{A spectral fit to 2A0335, a large-absorption column cluster.  The
quantities shown are the same as in Fig. \ref{Fig_A1795fit}.
\label{Fig_2A0335fit}}
 
\vspace{15pt}
\noindent Fig\_A1795con.ps

\figcaption{$\Delta \chi$-square contours in normalization vs. $\NHxo$ for
A1795.  Moving outward, the contours have values of 2.3, 4.6, and 9.2,
corresponding to confidence limits of 68\%, 90\%, and 99\%.
\label{Fig_A1795con}}

\vspace{15pt}
\noindent Fig\_2A0335con.ps

\figcaption{$\Delta \chi$-square contours in normalization vs. $\NHxo$ for
2A0335.  The contours have the same values as those in Fig. \ref{Fig_A1795con}.
\label{Fig_2A0335con}}
 
\vspace{15pt}
\noindent Fig\_NHxNH.ps

\figcaption{The ratio of X-ray column density to the 21 cm column of
Hartmann and Burton (1997).  The lines of sight are toward the 20 sample
clusters.  The open circles represent directions of low column density
which are used to estimate a fiducial $\NHxo/\NH21$.  \label{Fig_NHxNH}}

\vspace{15pt}
\noindent Fig\_IRAS1.jpg \\
\noindent Fig\_IRAS2.jpg

\figcaption{\IRAS\ \mic100\ images toward each of the clusters in the
sample.  Each frame is 7.5$^{\circ}$ on a side, and the location of the
cluster is indicated with a 40$^{\prime}$ diameter circle.
\label{Fig_IRAS}}

\vspace{15pt}
\noindent Fig\_HHe.ps

\figcaption{A comparison of X-ray column densities for the low column density
sight lines for two values of He/H.  The error bars do not include
uncertainties in $\NHoo$.
\label{Fig_HHe}}
 
\vspace{15pt}
\noindent Fig\_H2frac.ps

\figcaption{The fractional \H2\ column as a function of the total inferred
H column N(HI+\H2).  The triangles represent \H2\ fractions derived by
Savage \etal\ (1977) from ultraviolent absorption of molecular hydrogen.
\label{Fig_H2frac}}
 
\vspace{15pt}
\noindent Fig\_ann.gif

\figcaption{Differential (dotted) and cumulative (solid) averages of the
\IRAS\ \mic100\ flux in the direction of the seven high-column clusters.
The quantity $\delta f_{\nu}$ is defined as the relative difference between
$\langle$\F100$\rangle$ averaged over the entire 36$^{\prime}$ wide disk
and $\langle$\F100$\rangle$ averaged over the 6-12$^{\prime}$ annulus,
corresponding to the HB beam and a fiducial X-ray column measurement.
\label{Fig_ann}}
 
\vspace{15pt}
\noindent Fig\_NHF100.gif

\figcaption{Hydrogen column density determined from X-ray absorption (A) and
from 21 cm emission (B; Hartmann and Burton 1997) versus \IRDI\ \mic100\ flux
(Schlegel \etal\ 1998) for the 20 clusters in the sample. \label{Fig_NHF100}}

\vspace{15pt}
\noindent Fig\_IR21.ps

\figcaption{21 cm column (Hartmann and Burton 1997) versus \IRDI\ \mic100\
flux (Schlegel \etal\ 1998) for the 13 low-column clusters in the sample.
At $\NHIo=0$, the \mic100\ flux is $0.04 \pm 0.31$ MJy sr$^{-1}$,
consistent with zero.
\label{Fig_IR21}}

\vspace{15pt}
\noindent Fig\_IRX.ps

\figcaption{X-ray absorption column versus \IRDI\ \mic100\ flux (Schlegel
\etal\ 1998) for the 20 clusters in the sample.  At $\NHIo=0$, the \mic100\
flux is $0.26 \pm 0.10$ MJy sr$^{-1}$.  If we use only the 13 low-column
clusters, this flux becomes $0.22 \pm 0.24$ MJy sr$^{-1}$.  The inset is
an enlargement of the region near the origin.
\label{Fig_IRX}}

\end{document}